\documentclass[aps,prl,reprint,amsmath,amssymb,floatfix]{revtex4-1}

\usepackage{bm,graphicx,amsmath,gensymb,siunitx}
\usepackage{hyperref}

\hypersetup{
	colorlinks=true,
	linkcolor=blue,
	citecolor=blue,
	filecolor=magenta,
	urlcolor=cyan
}

\graphicspath{
{Figures/}
{../Figures/}
}

\allowdisplaybreaks[1]

\newcommand{\la}{\langle}
\newcommand{\ra}{\rangle}
\newcommand{\kk}{{\bm{k}}}
\newcommand{\GG}{\bm{G}}
\newcommand{\rr}{\bm{r}}
\newcommand{\RR}{\bm{R}}
\newcommand{\ttau}{\bm{\tau}}
\newcommand{\hpp}{\hat{\bm{p}}}
\newcommand{\hp}{\hat{p}}
\newcommand{\hrr}{\hat{\bm{r}}}
\newcommand{\hx}{\hat{x}}
\newcommand{\hH}{\hat{H}}
\newcommand{\hV}{\hat{V}}
\newcommand{\hP}{\hat{P}}
\newcommand{\tkk}{\tilde{\bm{k}}}
\newcommand{\tGG}{\tilde{\bm{G}}}
\newcommand{\tE}{\tilde{E}}
\newcommand{\tX}{\tilde{X}}
\newcommand{\tn}{\tilde{n}}
\newcommand{\tH}{\tilde{H}}
\newcommand{\e}{\text{e}}
\newcommand{\ii}{\text{i}}

\newcommand{\cS}{\mathcal{S}}

\newcommand{\papertitle}{Theory of quantum oscillations in quasicrystals: Quantizing spiral Fermi surfaces}

\newcommand{\tcm}{T.C.M. Group, Cavendish Laboratory, University of Cambridge, JJ Thomson Avenue, Cambridge CB3 0HE, United Kingdom}

\begin{document}
	
\title{\papertitle}

\author{Stephen Spurrier}
\author{Nigel R. Cooper}

\affiliation{\tcm}

\date{\today}


\begin{abstract}
We show that electronic materials with disallowed rotational symmetries that enforce quasiperiodic order can exhibit quantum oscillations and that these are generically associated with exotic ``spiral Fermi surfaces." These Fermi surfaces are self-intersecting, and characterized by a winding number of their surface tangent---a topological invariant---that is larger than one. 
We compute the nature of the quantum oscillations in two experimentally relevant settings which give rise to spiral Fermi surfaces: a ``nearly-free-electron" quasicrystal, and \ang{30} twisted bilayer graphene.
\end{abstract}

\maketitle


{\emph{Introduction}.} Quasiperiodic systems are long-range ordered and yet nonperiodic~\cite{steinhardt1987physics,trebin2006quasicrystals}. This places them in a fascinating intermediate regime between periodic and disordered~\cite{sokoloff85unusual,poon92electronic}. They first entered physics with the discovery by Shechtman \emph{et al.}~\cite{shechtman84metallic} of quasicrystals---electronic materials with crystallographically disallowed rotational symmetries~\cite{levine84quasicrystals}. 
However, due to the lack of Bloch's theorem, there still remain many open questions about their electronic properties. More recently, a surge of interest has risen due to the possibility of studying these questions in new, highly controllable, contexts such as cold atoms~\cite{guidoni1999atomic,viebahn2019matter} and photonics~\cite{chan1998photonic}, allowing for the exploration of physics such as localization~\cite{fallani2007ultracold,schreiber15observation,khemani2017two}, topology~\cite{kraus12topologicalstates,bandres16topological,fulga2016aperiodic,tran2015topological,huang2018quantum,dareau2017revealing,dana2014topologically}, and synthetic dimensions~\cite{viebahn2019matter,kraus13fourdimensional}.

While progress is being made in artificial quasiperiodic systems, new avenues are also opening for electronic quasicrystals, including a recent realization of quasicrystalline \ang{30} twisted bilayer graphene~\cite{ahn2018dirac,yao2018quasicrystalline,lin2018surfactant}. One of the key tools for studying periodic electronic materials are quantum oscillations~\cite{shoenberg2009magnetic,dehaas1930dependence,lifshitz1956theory}---a well-established technique for characterising Fermi surfaces based on the semiclassical quantisation of orbits into Landau levels~\cite{onsager1952interpretation,xiao10berry}. For quasicrystals, one might expect that quantum oscillations are precluded by the lack of a well defined Fermi surface or the typically low conductivity~\cite{poon92electronic}. Nevertheless, an early experimental study surprisingly found these to present~\cite{haanappel1998dehaas}. Despite this finding, there has yet to be a theory developed to explain how quantum oscillations could occur in a quasicrystal~\cite{roche1998fermi,krajci2001fermi}.

In this Rapid Communication, we develop such a theory. We show how quantum oscillations can occur in quasicrystals, using two experimentally relevant models as examples: a nearly-free-electron quasicrystal~\cite{rotenberg2000quasicrystalline,theis03electronic,rogalev2015fermi,smith87pseudopotentials,carlsson1993bandgap} and \ang{30} twisted bilayer graphene~\cite{yao2018quasicrystalline,ahn2018dirac,lin2018surfactant}. Surprisingly, we find that when quantum oscillations do occur, these are associated with an unconventional type of Fermi surface---which we dub a ``spiral Fermi surface''---with topological character. Moreover, we find that the presence of a spiral Fermi surface in quasicrystals is generic---the only requirement is a separation in energy scales of their gaps~\cite{SI,gaps}.

\begin{figure}[tbp]
	\centering
	\includegraphics[trim={.0cm .cm .0cm .0cm}, clip,width=.99\linewidth]{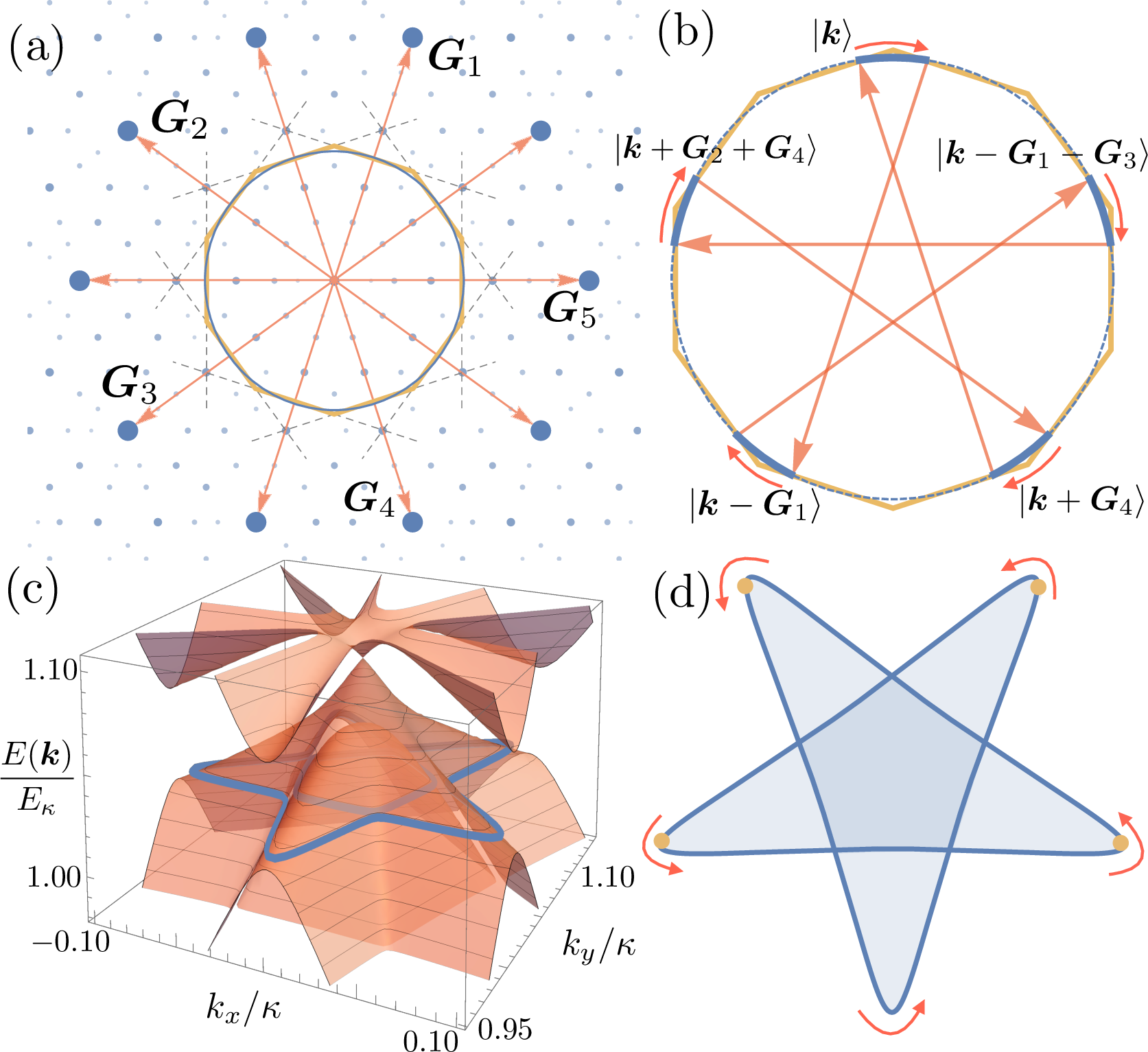}%
	\caption{\label{fig:QONFE}\textbf{Quantum oscillations in a nearly-free-electron quasicrystal}. (\textbf{a}) Sketch of our model for a nearly-free-electron quasicrystal, consisting of ten Fourier components (largest blue points) at momenta $ \pm \GG_i $ (red arrows) which have Bragg planes (gray dotted lines) forming a pseudo-Brillouin-zone (orange decagon) and which intersect the free-electron Fermi surface (blue circle). 
	All combinations of these 10---the reciprocal lattice of periodic systems---covers $k$-space densely (smaller blue points).
	(\textbf{b}) The semiclassical trajectories in an external magnetic field (solid blue curves) drift along the free electron Fermi surface (dotted circle) with scattering by each $ \GG_i $ at the pseudo-Brillouin-zone boundary. 
	(\textbf{c}) These semiclassical trajectories are seen as constant energy contours (blue curves) of an ``effective band structure,'' which is shown for $ V_0 = 0.1 E_\kappa $.
	(\textbf{d}) The resulting spiral Fermi surface with nontrivial turning number of $ N_\text{t}  = 2 $.
	}
\end{figure}

The topology of the spiral Fermi surfaces is classified using the turning number $ N_\text{t} $, which is defined as the winding number of the surface tangent---an invariant for two-dimensional plane curves~\cite{whitney1937regular,bovenzi2018twisted}. A Fermi surface that can be smoothly deformed to a circle has $ N_\text{t} = \pm 1 $ and is considered trivial, while all other turning numbers are considered nontrivial.
In quasicrystals, nontrivial turning numbers generically occur when the Fermi surface winds the pseudo-Brillouin-zone corner~\cite{SI}.
As such, the presence of nontrivial turning numbers is related to their crystallographically disallowed $ 2m $-fold rotational symmetry according to~\cite{SI} (with quasicrystals satisfying $ m \ge 4 $)
\begin{align}
N_\text{t} = \begin{cases}
m-1 \quad &\text{if}\ m\ \text{even}, \\
(m-1)/2 \quad &\text{if}\ m\ \text{odd}.
\end{cases}\label{eq:TurningNoRotational}
\end{align} 
Moreover, the turning number is measurable experimentally in quantum oscillations. This appears in the so-called Maslov contribution~\cite{keller1958corrected,arnold1967characteristic,ozorio1990hamiltonian,alexandradinata2018semiclassical,xiao10berry} to the offset $ \gamma $ in semiclassical quantization~\cite{onsager1952interpretation,lifshitz1956theory,shoenberg2009magnetic,xiao10berry},
\begin{align}
\ell_B^2 S(E) = 2 \pi (n+\gamma), \label{eq:onsager}
\end{align}
where $ \ell_B = \sqrt{\hbar/eB} $ is the magnetic length, $ S(E) $ is the area swept out by the wave packet in reciprocal space, $ n $ is an integer, and
\begin{align}
\gamma &= \frac{N_\text{t}}{2} - \frac{\varphi_\text{Berry}}{2\pi}.
\end{align}
The first term of the right hand side is the topological Maslov contribution~\cite{keller1958corrected,arnold1967characteristic,ozorio1990hamiltonian,alexandradinata2018semiclassical,xiao10berry}, while the second part is the geometrical Berry phase contribution~\cite{mikitik1999manifestation,xiao10berry}. 


{\emph{Nearly-free-electron quasicrystal}.} Our first model is for a nearly-free-electron quasicrystal with an axis of tenfold rotational symmetry. This model is an approximation to icosahedral~\cite{shechtman84metallic} and decagonal quasicrystals~\cite{bendersky1985quasicrystal} that have nearly-free-electron qualities~\cite{friedel1988metallic,vaks1988possible,fujiwara1991universal,poon92electronic}, such as the various aluminum-based quasicrystals studied in angle-resolved photoemission spectroscopy~\cite{rotenberg2000quasicrystalline,theis03electronic,rogalev2015fermi}, and is known to exhibit a spiral holonomy~\cite{spurrier2017semiclassical}. We consider the two-dimensional single-particle spinless Hamiltonian $ \hat{H} = \hpp^2/2m + V(\hrr) $, with potential in real space given by~\cite{smith87pseudopotentials,lu1987electronic,carlsson1993bandgap}
\begin{align}
V(\rr) \equiv 2 V_0 \sum_{j=1}^{5} \cos \GG_j \cdot \rr,
\end{align}
where $ \GG_j \equiv 2\kappa \left(\cos  2\pi j/5, \sin 2\pi j/5 \right) $ and $ V_0 $ is the strength of each individual Fourier component, which is assumed to satisfy the nearly-free-electron limit $ V_0 \ll E_\kappa \equiv \hbar^2\kappa^2/2m $. This model amounts to keeping the ten dominant Fourier components (i.e.,\ the brightest spots in the diffraction pattern), in particular, those with Bragg planes that intersect the free-electron Fermi surface.

Our analysis of this model relies on the nearly-free-electron limit. This tells us that the free-electron Fermi surface [the blue circle shown in Fig.~\ref{fig:QONFE}(a)] remains mostly unchanged except for the opening of gaps proportional to $ V_0 $ at intersections with Bragg planes to $ \pm\GG_i $ [dotted lines in Fig.~\ref{fig:QONFE}(a)] and also gaps from combinations of $ n $ Bragg reflections. Since all combinations of $ \GG_i $ form a dense set in $ k $ space [shown in Fig.~\ref{fig:QONFE}(a)], the set of all associated gaps will also be dense. Crucially, these gaps form a distinct hierarchy, $ \Delta_\text{gap}^{n^\text{th}} \propto (V_0/E_\kappa)^n $. Thus, for $V_0/E_\kappa$ small, one can choose a magnetic field that removes $ (n+1) $th-order gaps via magnetic breakdown, while keeping $ n $th-order gaps. The probability of magnetic breakdown is given by $ P_\text{MB} = \e^{-\pi ab \ell_B^2} $,
where $ a $ and $ b $ are the axes of the avoided crossing hyperbola~\cite{alexandradinata2018semiclassical,cohen1961magnetic,pippard1962quantization,blount1962bloch,reitz1964magnetic}.
The simplest scenario is the regime of fields in which only first-order gaps are kept,
\begin{align}
\left(\frac{V_0}{E_\kappa}\right)^{4} \ll \frac{\hbar \omega_c}{E_\kappa} \ll \left(\frac{V_0}{E_\kappa}\right)^{2}, \label{eq:semiadiabatic}
\end{align}
where $ \omega_c \equiv e B/m $ is the cyclotron frequency. We refer to this as the ``first-order regime'' of fields. The relevant gaps in this regime are along the pseudo-Brillouin-zone edges [yellow decagon in Fig.~\ref{fig:QONFE}(b)].

Having specified an appropriate regime of magnetic fields---the first-order regime---the semiclassical trajectories can be found by tracing a path along the unperturbed free-electron Fermi surface and making jumps at intersections with relevant Bragg planes. This procedure is shown in Fig.~\ref{fig:QONFE}(b) for a wavepacket that is initially localized at the top of the pseudo-Brillouin zone in the free-particle state $ |\kk\ra $. This state proceeds clockwise around the free-electron Fermi surface until it encounters the Bragg plane to $ \GG_1 $, at which point it is scattered into the state $ |\kk-\GG_1\ra $. Continuing in this manner the wave packet is scattered a total of five times between the following states,
\begin{align}
|\kk\ra \to& |\kk-\GG_1\ra \to |\kk - \GG_1-\GG_3\ra \nonumber\\ 
&\to |\kk+\GG_2+\GG_4\ra \to |\kk +\GG_4\ra \to |\kk\ra, \label{eq:subset}
\end{align}
after which the wave packet returns and can be quantized according to \eqref{eq:onsager}. 

By projecting onto the above subset of states we find an effective band structure, as shown in Fig.~\ref{fig:QONFE}(c).  
The semiclassical trajectories described qualitatively above [blue curves in Fig.~\ref{fig:QONFE}(b)] can now be seen quantitatively as the constant energy contours of this band structure shown in Figs.~\ref{fig:QONFE}(c) and (d). The turning number can be computed by using a sum over the extremal points (points with vertical tangent), $ N_\text{t} = \frac{1}{2}\sum_i \nu_i $, where $ \nu_i = \pm1 $ for an extremal point with anticlockwise (clockwise) orientation. For the Fermi surface in Fig.~\ref{fig:QONFE}(d), there are four extremal points (yellow dots) with anticlockwise orientation. This gives a turning number of $ N_\text{t} = 2 $, as expected from \eqref{eq:TurningNoRotational} with $ m=5 $. This spiral Fermi surface (with $ N_\text{t} = 2 $) does not require fine tuning of the Fermi energy, unlike the ``twisted Fermi surface'' (with $ N_\text{t} = 0 $) of a tilted Weyl point~\cite{bovenzi2018twisted}.

\begin{figure}[t]
	\centering
	\includegraphics[trim={.1cm .5cm 0.1cm .1cm}, clip,width=.99\linewidth]{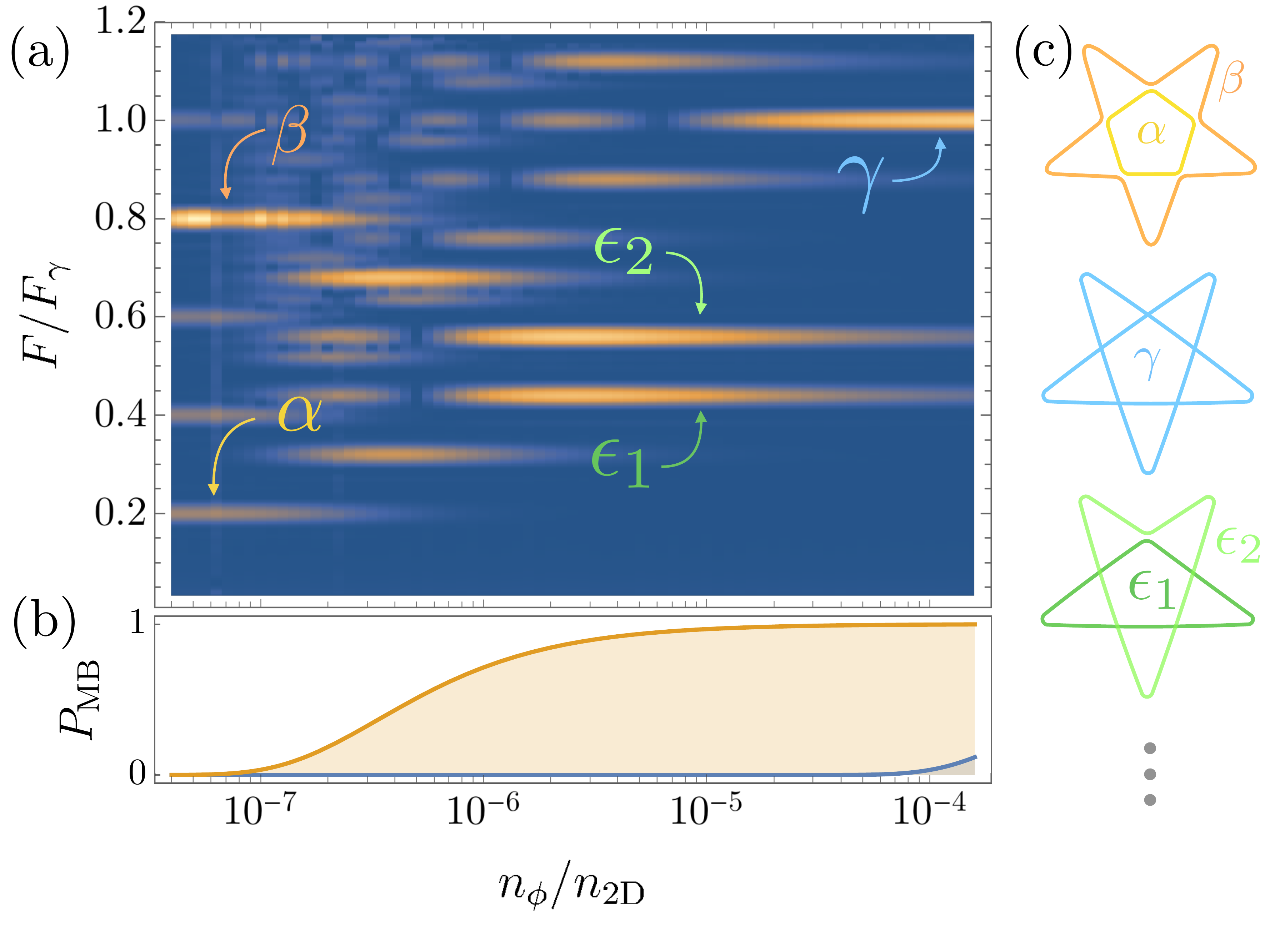}%
	\caption{\label{fig:MBNFE}\textbf{Quantum oscillation frequency spectrum across the transition between second order and first order field regimes}. (\textbf{a}) Density plot of the frequency spectrum (with frequencies given as a ratio of the spiral Fermi surface frequency $ F_\gamma $) as a function of magnetic field (given in terms of $ n_\phi = eB/h $ and $ n_\text{2D} = k_\text{F}^2/2\pi $), for the parameters $ V_0/E_\kappa = 0.01 $ and $ E_\text{F}/E_\kappa = 1.01$. A transition is seen from a single frequency ($ \gamma $) at high fields ($ n_\phi/n_\text{2D} \approx 10^{-4} $) to a pair of frequencies ($ \alpha $ and $ \beta $) at lower fields ($ n_\phi/n_\text{2D} \approx 10^{-7} $). (\textbf{b}) Plot of the magnetic breakdown probabilities at second-order gaps (orange curve) and first-order gaps (blue curve). (\textbf{c}) A selection of semiclassical contours used to label frequencies. For intermediate fields a complex mix of frequencies is present that can be labeled using intermediate contours such as $ \epsilon_1 $ and $ \epsilon_2 $.
	}
\end{figure}

We highlight two key signatures of this nontrivial turning number for quantum oscillations. The first is for the offset $ \gamma $ in the semiclassical quantization \eqref{eq:onsager}. A conventional Fermi surface that is deformable to a circle results in $ \gamma = 1/2 $, with deviations from this indicating a nonzero Berry phase. Here, $ N_\text{t} = 2 $ results in $ \gamma = 0 $ for \emph{zero} Berry phase. The second signature is related to the ``magnetic breakdown transition'' as a function of magnetic field between first-order and second-order field regimes, shown in Fig.~\ref{fig:MBNFE}. As this transition occurs at a fixed Fermi energy, the total turning number is conserved~\cite{SI}. Since the transition is between an odd number of dominant frequencies (a single frequency $ \gamma $) and an even number ($ \alpha $ and $ \beta $), at least one Fermi surface must be nontrivial~\cite{SI}.


In order to address these results experimentally, one must consider three key parameters: the Fermi energy $ E_\text{F} $, the field $ B $, and the potential $ V_0 $. In typical nearly-free-electron quasicrystals, $ E_\text{F} $ is already at the required location---with the free-electron Fermi surface intersecting the pseudo-Brillouin-zone boundary \footnote{This well known condition is closely connected to the stability of the quasicrystalline phase, often referred to as the Hume-Rothery rule~\cite{poon92electronic,bancel1986icosahedral,smith87pseudopotentials,hafner1992electronic,martin2016weak}.}---therefore little to no doping should be required. For the model parameters used in Fig.~\ref{fig:MBNFE}, the flux density required to reach the first-order regime is small compared to the electron density. Using typical experimental parameters of $ \kappa = \SI{1.3}{\per\angstrom} $  and $ m = m_\text{e} $ (the free-electron mass)~\cite{theis03electronic}, this occurs for fields of $ B \approx \SI{10}{\tesla} $---a regime attainable experimentally. The required potential $ V_0 $, however, provides the most severe constraint experimentally. The calculation of the magnetic breakdown transition in Fig.~\ref{fig:MBNFE} allows us to quantify the maximum allowed $ V_0/E_\kappa $---for a ratio that is too large the two regimes (first and second order) are not distinguishable. Using these criteria we find a maximum of $ V_0/E_\kappa \approx 0.02 $, which corresponds to a gap at the pseudo-Brillouin-zone edge of approximately \SI{0.2}{\electronvolt}. Additionally, a small ratio of $ V_0/E_\kappa $ ensures the pseudogap at $ E_\text{F} $ is not fully formed~\cite{carlsson1993bandgap}, and the system remains metallic.

{\emph{Twisted bilayer graphene}.} Our second model is for \ang{30} twisted bilayer graphene, a system that has recently seen its first experimental realization~\cite{ahn2018dirac,yao2018quasicrystalline,lin2018surfactant}. This incommensurate superstructure satisfies the typical definition of a quasicrystal~\cite{steinhardt1987physics} in that its diffraction pattern contains sharp peaks possessing a $ 2m $-fold symmetry (here $ m  = 6 $) that requires more basis vectors (four) than dimensions (two) in order to be indexed~\footnote{It does not satisfy other definitions of a quasicrystal~\cite{yu2018comment}.}.
The quasiperiodic structure of the diffraction peaks is sufficient to cause the effective band structure to exhibit a spiral Fermi surface with a highly nontrivial turning number of $ N_\text{t}=5 $. 

To show this, we use the model of twisted bilayer graphene developed in Ref.\@ \onlinecite{koshino2015interlayer}.
This takes a standard nearest-neighbor tight-binding Hamiltonian $ H_\parallel $ for each layer, which is off diagonal in a Bloch basis $ |\kk,X\ra $, with $ X = A, B $ sublattice indices~\cite{castro2009electronic},
\begin{align}
\la \kk , A | H_\parallel | \kk, B \ra  = -t \sum_{i=1}^{3} \e^{-i \kk \cdot \bm{\rho}_i},
\end{align}
where the vectors $ \bm{\rho}_i $ connecting nearest neighbors in layer 1 are rotated by \ang{30} with respect to those in layer 2. Tunneling between the layers causes a Bloch state from layer 1 with crystal momentum $ \kk $ to be coupled to all those from layer 2 with crystal momentum $ \tkk = \kk +\GG - \tGG $~\cite{koshino2015interlayer},
\begin{align}
\la \tkk, \tX | H_\perp | \kk, X \ra = - t_\perp (\kk+\GG) \e^{-\ii\GG \cdot \ttau_X+\ii\tGG\cdot \ttau_{\tX}},
\end{align}
where a tilde (no tilde) denotes layer 2 (1), $ \GG $ is a reciprocal lattice vector, $ \ttau_X $ are position vectors of the sublattice sites within the unit cell, and $ t_\perp(\kk) $ is radially symmetric and decays exponentially for $ \kk $ beyond the first Brillouin zone~\cite{tblg1}.

\begin{figure}[t]
	\centering
	\includegraphics[trim={.0cm .0cm 0.cm .0cm}, clip,width=1.\linewidth]{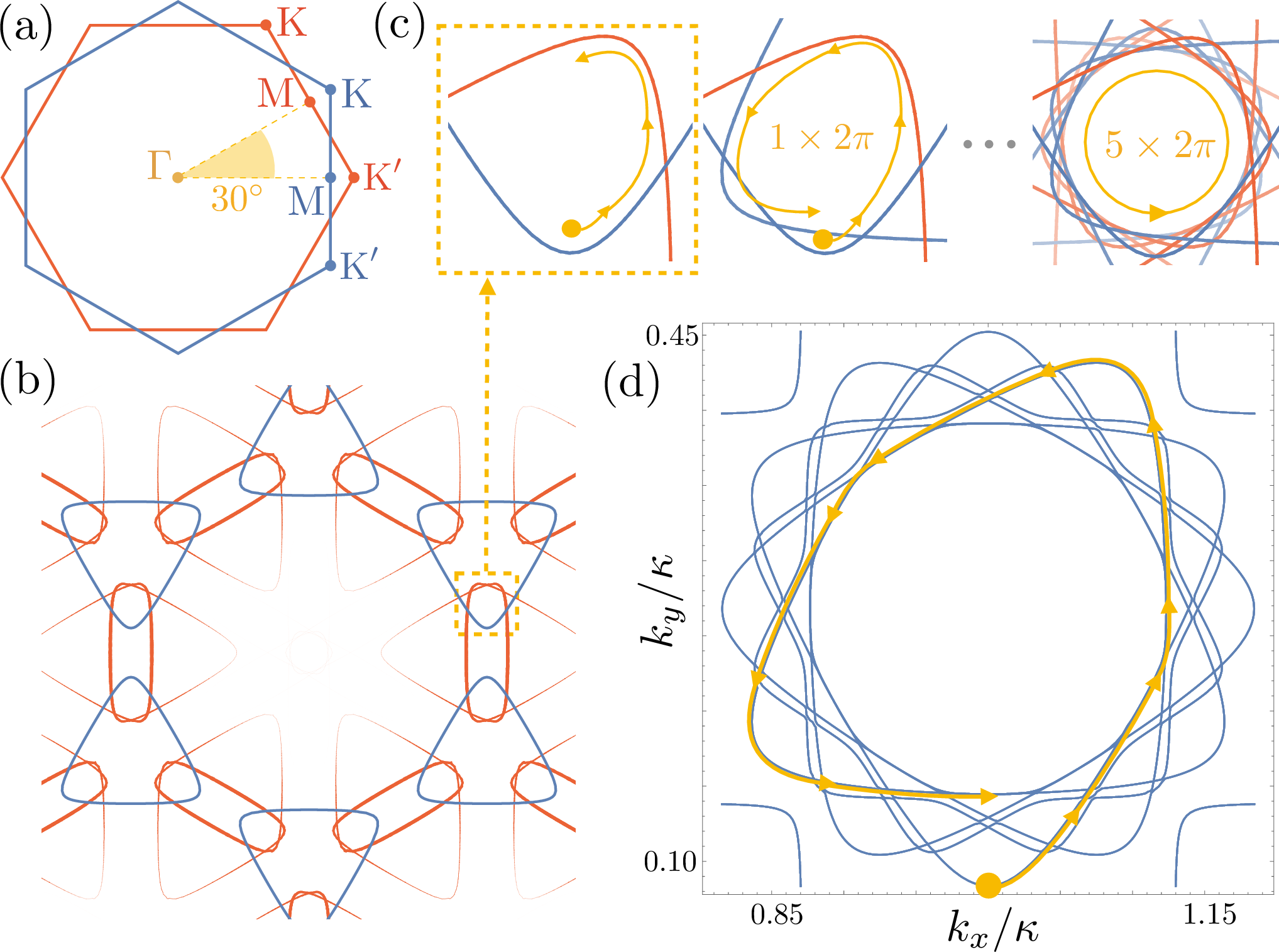}
	\caption{\label{fig:QOTBLG}\textbf{Quantum oscillations in incommensurate 30$\degree$ twisted bilayer graphene}. (\textbf{a}) Sketch of each layer's Brillouin-zone, blue is referred to as layer 1 and red as layer 2. (\textbf{b}) The Fermi surface of layer 1 is coupled to the Fermi surface of layer 2, in addition to all possible translations of this Fermi surface by reciprocal lattice vectors from layer 1. (\textbf{c}) For sufficiently large doping, the Fermi surface of layer 1 intersects that of layer 2, allowing a semiclassical trajectory that jumps from layer 1 to layer 2. This repeats a total of 12 times before returning to be quantised by semiclassical quantisation. (\textbf{d}) The Fermi surface of an effective model, shown for the experimental parameters given in Ref.~\onlinecite{tblg2}.
	}
\end{figure}

We analyze this model by assuming a weak coupling between the two layers, $ t_\perp(\kk) \ll t $. 
As with the nearly-free-electron limit in the previous section, this assumption is key to deriving meaningful semiclassical trajectories. In particular, this allows us to assert that the Fermi surfaces of each layer will be little affected, except at degenerate points that satisfy
\begin{align}
E(\kk) = \tE(\tkk), \quad \kk+\GG = \tkk +\tGG, \label{eq:graphenedegeneracy}
\end{align} 
where $ E(\kk) $ and $ \tE(\tkk) $ are the bandstructures of the unperturbed layers 1 and 2. This is considered a first order coupling, as a gap will open proportional to  $ t_\perp $. However there will also be gaps opened due to second order processes that couple a layer to itself at the following degeneracies,
\begin{align}
E(\kk) = E(\kk+\tGG), \quad \tE(\kk) = \tE(\kk+\GG),
\end{align}
with these gaps proportional to $ t_\perp^2/t $. For simplicity we choose to work in a field regime in which second-order, intralayer gaps can be ignored, while interlayer are kept---which can be safely assumed to exist given the weak-coupling assumption. However, this still leaves a dense set of gaps given by \eqref{eq:graphenedegeneracy}. Fortunately, many of these are exponentially small due to the $ \kk $ dependence of $ t_\perp(\kk) $, as shown in Fig.~\ref{fig:QOTBLG}(b). We therefore choose a field such that those gaps opened by the exponential tail of $ t_\perp(\kk) $ beyond the first Brillouin zone are ignored. In this regime, one finds that spiral Fermi surfaces appear at those dopings for which the original Fermi surfaces of the two layers intersect.
 
Having determined a suitable field regime and doping, we derive the semiclassical trajectories by simply tracing a path along the unperturbed Fermi surfaces and switching between layers at the relevant intersections. If one begins this process, as shown in Fig.~\ref{fig:QOTBLG}(c), on layer 1 (blue contour), the wave packet will progress anticlockwise before jumping to layer 2 (red contour). Here, it essentially repeats this contour again, now rotated by $ 5\pi/6 $. This occurs a total of 12 times, resulting in a trajectory that winds the center a total of five times, as shown in Fig.~\ref{fig:QOTBLG}(c). This is reflected in the effective Fermi surface shown in Fig.~\ref{fig:QOTBLG}(d), shown for typical model parameters. The turning number is computed by summing over the extremal points (as discussed in the previous section). For each $ 2\pi $ winding about the center there are two extremal points with anticlockwise orientation. As the Fermi surface winds the center five times, there are a total of ten extremal points with anticlockwise orientation, which means $ N_\text{t} = 5 $, as expected from \eqref{eq:TurningNoRotational} with $ m=6 $.

\begin{figure}[t]
	\centering
	\includegraphics[trim={.0cm 0cm 0cm 2.6cm}, clip,width=.75\linewidth]{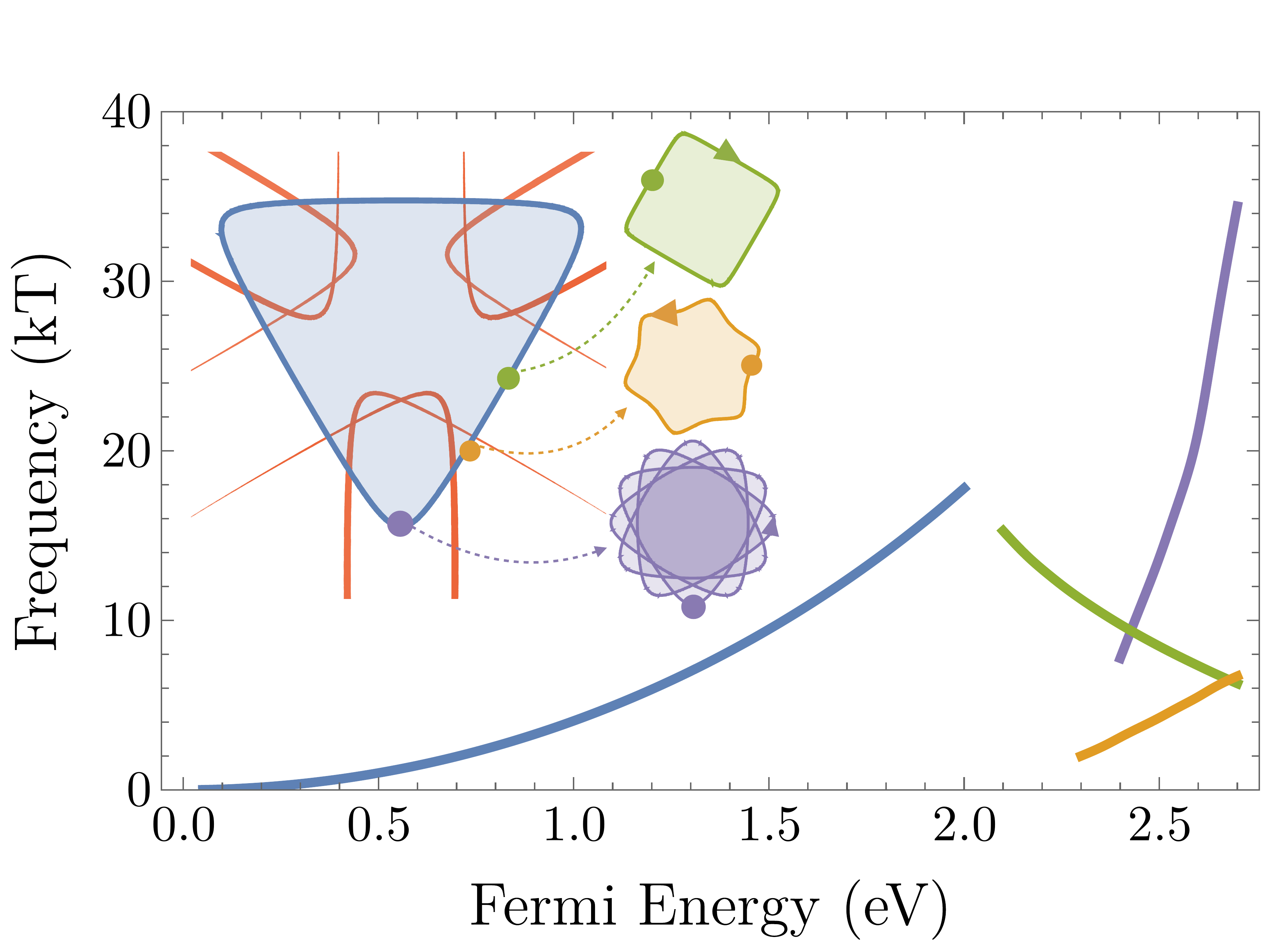}%
	\caption{\label{fig:FreqTBLG}\textbf{Phenomenology for 30\degree~twisted bilayer graphene}. 
	Plot of quantum oscillation frequencies as a function of doping away from charge neutrality, using the same parameters given in Fig.~\ref{fig:QOTBLG}. \textbf{Inset}: Each frequency is identified with a different starting point on the unperturbed ``Dirac'' Fermi surface. The purple contours are those with a nontrivial turning number and have a larger than naively expected frequency.
	}
\end{figure}

Since the turning number in this case is odd, the two signatures highlighted in the previous section for a nearly-free-electron quasicrystal do not apply here---$ \gamma = 1/2 $,   which is indistinguishable from the trivial case and breakdown transitions cannot identify odd turning numbers~\cite{SI}. Instead, the key signature here is in the dependence of the quantum oscillation frequency on doping, as shown in Fig.~\ref{fig:FreqTBLG}. Below a critical doping of approximately \SI{2}{\electronvolt}, the Fermi surfaces of each layer do not cross and only a single frequency is present, associated with a broadening Dirac cone. However, above \SI{2}{\electronvolt} the two layers' Fermi surfaces intersect and form three distinct contours: a square holelike contour that is topologically trivial ($ N_\text{t} = -1 $), a hexagonal electronlike contour that is also trivial ($ N_\text{t} = 1 $), and finally the nontrivial ($ N_\text{t}= 5 $) electron-like contour derived above. Crucially, the spiral Fermi surface is distinguished by a sharp increase of frequency with doping due to multiple overlapping Fermi surface areas. 

We address the experimental feasibility by using tight-binding parameters for bilayer graphene~\cite{tblg2}. This leaves two key parameters: the required field strength $ B $ and doping $ E_\text{F} $. By extracting gap parameters we determine the required field to be $ B \approx \SI{7}{\tesla} $, which is within experimental capabilities. The key challenge experimentally will be to reach a doping of $ E_\text{F} \approx \SI{2}{\electronvolt} $, although larger dopings have been realized experimentally for single layers~\cite{mcchesney2010extended}. 

In summary, we have used two very different models to show that quantum oscillations can arise in quasicrystalline materials. In both cases these are associated with spiral Fermi surfaces. In fact, this is a generic result, relying only on their unconventional rotational symmetry. As detailed in the Supplemental Material~\cite{SI}, this alone can be used to deduce the nontrivial turning number of the spiral Fermi surface. Moreover, that this arises for twisted bilayer graphene offers exciting opportunities for experimental observation.

In compliance with EPSRC policy framework on research data, all data are directly available within the publication.

\acknowledgements{
{\emph{Acknowledgments}.}~We gratefully acknowledge support from EPSRC via Grants No.~EP/K030094/1, No.~EP/P034616/1, and No.~EP/P009565/1, and from a Simons Investigator award of the Simons Foundation. 
}

\bibliography{../BibliographyQO.bib}

\begin{thebibliography}{78}
\bibitem [{gap()}]{gaps}%
  \BibitemOpen
  \bibinfo {note} {Throughout we use the word `gap' in reference to an avoided
  crossing between two energy levels---not in reference to a gap in the energy
  spectrum.}\BibitemShut {Stop}%
\bibitem [{pha()}]{phases}%
  \BibitemOpen
  \bibinfo {note} {Note that the phases we use differ from those of Refs.
  \onlinecite{slutskin1968dynamics,alexandradinata2018semiclassical}. They are
  chosen here such that the theory produces the expected results in the two
  limits of $ \rho \to 0,\,1 $.}\BibitemShut {Stop}%
\bibitem [{apo()}]{apodization}%
  \BibitemOpen
  \bibinfo {note} {We process our data using an apodization, specifically, a
  Kaiser window with parameter $ \alpha = 1.7 $. This smoothly tapers the data
  to zero at the boundaries, suppressing unwanted ``artifacts'' that occur due
  to an otherwise sharp truncation.}\BibitemShut {Stop}%
\end{thebibliography}


\newpage

\clearpage

\appendix

\setcounter{figure}{0}
\makeatletter 
\renewcommand{\thefigure}{S\arabic{figure}}

\newcounter{defcounter}
\setcounter{defcounter}{0}

\newenvironment{myequation}
{%
	\addtocounter{equation}{-1}
	\refstepcounter{defcounter}
	\renewcommand\theequation{S\thedefcounter}
	\align
}
{%
	\endalign
}

\begin{onecolumngrid}
	\begin{center}
		{\fontsize{12}{12}\selectfont
			\textbf{Supplemental Material for ``\papertitle''\\[5mm]}}
		{\normalsize Stephen Spurrier and Nigel R. Cooper\\[1mm]}
		{\fontsize{9}{9}\selectfont  
			\textit{\tcm}}
	\end{center}
	\normalsize
\end{onecolumngrid}

\vspace{20pt}

\begin{twocolumngrid}

	\section{Generalities}
	
	In the main text we provide two simple experimentally relevant examples of quasicrystals for which quantum oscillations are both present and feature a spiral Fermi surface. Here we generalise these results by outlining the basic ingredients and the single necessary assumption---\emph{a separation in the gaps energy scales}---in order to find both properties in other quasicrystals. 
	
	{\emph{Quantum Oscillations}.---} We start by explaining why quantum oscillations are not present without this assumption: All quasicrystals have a densely gapped spectrum, with the location of each gap in a one-to-one correspondence with the dense set of reciprocal lattice vectors (momentum transfers). The possibility of finding single band adiabatic dynamics is then prohibited. Instead a state initialised in an eigenstate will generally diffuse across many eigenstates in presence of an external drive---thereby washing out any clear signal from the quantum oscillations. 
	
	On the other hand, if the assumption of a separation in the gaps energy scales is satisfied, quantum oscillations become possible. The two examples in the main text satisfy this assumption: nearly-free-electron and weakly coupled layers, while a third example is incommensurate charge density waves~\cite{zhang15disruption}. Under this assumption, while tunnelling between eigenstates is still unavoidable, the separation in energy scales allows one to choose an external drive that divides the gaps into two sets: those that are effectively removed by tunnelling and those that are respected by the adiabatic theorem and so are kept.

	{\emph{Spiral Fermi Surface}.---} While these basic arguments explain why quantum oscillations are present---and apply to any quasiperiodic system, not just quasicrystals---it is the \emph{disallowed rotational symmetries} of quasicrystals that result in the spiral Fermi surface: By selecting a subset of gaps that are kept, one simultaneously selects a subset of momentum transfers. These define a pseudo-Brillouin-zone (PBZ)~\cite{smith87pseudopotentials,lu88acoustic,burkov92optical}, which takes the form of a $ 2m $-sided regular polygon---note that for $ m = 2,3 $ a regular Brillouin zone is obtained (we will therefore treat both regular and quasicrystalline cases on an equal footing). Any semiclassical orbit that leaves this region is Bragg scattered by a momentum transfer returning it to the PBZ. Therefore all semiclassical orbits can be considered as trajectories within the PBZ. 
	
	Once the PBZ is justified, the possible turning numbers of orbits are readily classified under the constraint that in order to produce quantum oscillations these must be contractable~\cite{shoenberg2009magnetic}---for periodic lattices this would mean they should not wrap the Brillouin zone torus. Since we wish to emphasise the difference in this classification between a PBZ and a regular BZ, we add the additional constraint that these should not explicitly self-intersect on the PBZ/BZ. For the case of periodic lattices, $ m = 2,3 $, these conditions force all orbits to have $ N_\text{t} = \pm 1 $.

	The classification for the quasicrystalline case, $ m > 3 $, is carried out by first identifying a type of trajectory that is both contractable and is not self-intersecting, and yet has a nontrivial turning number, before going on to show why this is the only orbit that has these properties. 
	
	The trajectory we consider is one that circles a corner of the PBZ. This is the case for both examples given in the main text. The turning number here is simply calculated by finding the total angle swept out with respect to the corner for a complete loop. Furthermore, the results divide into two cases depending on the parity of $ m $, where $ m $ is defined above as half the number of sides of the PBZ: For even $ m $, the trajectory visits all $ 2m $ corners and sweeps out an angle of $ \pi - 2\pi/2m $ at each, giving a turning number for the even case of,
	\begin{equation}\label{turningeven}
	N_\text{t}^\text{even} = \frac{1}{2\pi} \left(\pi - \frac{2\pi}{2m}\right) 2m = m-1.
	\end{equation}
	For odd $ m $, the trajectory visits $ m $ corners (half of the total $ 2m $) and also sweeps out an angle of $ \pi - 2\pi/2m $ at each, giving a turning number for the odd case of,
	\begin{equation}\label{turningodd}
	N_\text{t}^\text{odd} = \frac{1}{2\pi} \left(\pi - \frac{2\pi}{2m}\right) m = \frac{m-1}{2}.
	\end{equation}
	For quasicrystalline symmetry, $ m > 3 $, the turning number for this trajectory is nontrivial, and yet is contractable (to the corner itself) and is not self-intersecting. 
	
	Note that while spiral Fermi surfaces necessarily self-intersect in an extended PBZ, since they have a turning number $ N_\text{t} \neq \pm1 $, they do not have to self-intersect when they are \emph{reduced to the PBZ}. This is seen in Fig.~\ref{fig:QONFE}b in the construction of the spiral Fermi surface for the NFE model and in Fig.~\ref{fig:torus}a for a PBZ with $ m = 4 $. 
	
	In order to show that this is the only trajectory that satisfies these properties, we appeal to the following construction: Since all edges of the PBZ are associated to their opposite, the PBZ is a closed manifold. However, unlike a BZ which has the topology of a torus, a PBZ has the topology of a \emph{higher genus torus}. We delay the details of this construction for now, instead focusing on how the corners of the PBZ map onto this higher genus manifold. For even $ m $, it can be shown that all $ 2m $ are identified, and therefore all $ 2m $ corners map to a single point on the closed manifold. A contractable loop that has no self-intersections can either enclose the point or not enclose it. If it encloses the point, the turning number will be nontrivial and given by the reasoning above. If it does not enclose the point, the turning number will be trivial. A very similar argument can be given for odd $ m $, however here $ m $ of the $ 2m $  corners map onto a single point while the other $ m $ corners map onto a different point. Therefore there is also the possibility of the loop encircling both points---if so, the turning number will also be nontrivial and will be given by $ N_\text{t}^\text{odd (both)} = m-2 $. 
	
	We have therefore classified the turning numbers of all contractible non-self-intersecting trajectories for $ m > 1 $, showing that nontrivial turning numbers occur only in the quasicrystalline case, $ m > 3 $. Moreover we see that the turning number is determined solely by the rotational symmetry of the PBZ.

	Finally, we provide a complementary explanation of these results that makes a close connection between the briefly mentioned higher genus topology of the PBZ and the special property of trajectories that enclose the corners. 
	
	To begin, we show how the genus is derived by computing the Euler characteristic, $ \chi $, which is given in terms of the number of vertices $ V $, edges $ E $ and faces $ F $~\cite{lee2006riemannian},
	\begin{equation}
	\chi = V-E+F.
	\end{equation}
	For even $ m $, the PBZ has 1 vertex, $ m $ edges and 1 face. While for odd $ m $, the it has 2 vertices, $ m $ edges and 1 face. This gives,
	\begin{align}
	\chi^\text{even} &= 2-m,  \\
	\chi^\text{odd} &= 3-m.
	\end{align}
	The genus, $ g $, is related to $ \chi $ (for orientable manifolds) by~\cite{lee2006riemannian},
	\begin{equation}
	\chi = 2-2g,
	\end{equation}
	which gives,
	\begin{align}
	g^\text{even} &= \frac{m}{2}, \\
	g^\text{odd} &= \frac{m-1}{2},
	\end{align}
	and therefore, for $ m > 3 $, we have $ g > 1 $, as stated above. An example is shown in Fig.~\ref{fig:torus}a-b for $ m=4 $.
	
		\begin{figure}[t]
		\centering
		\includegraphics[trim={.1cm 12.5cm 16.cm .2cm}, clip,width=.9\linewidth]{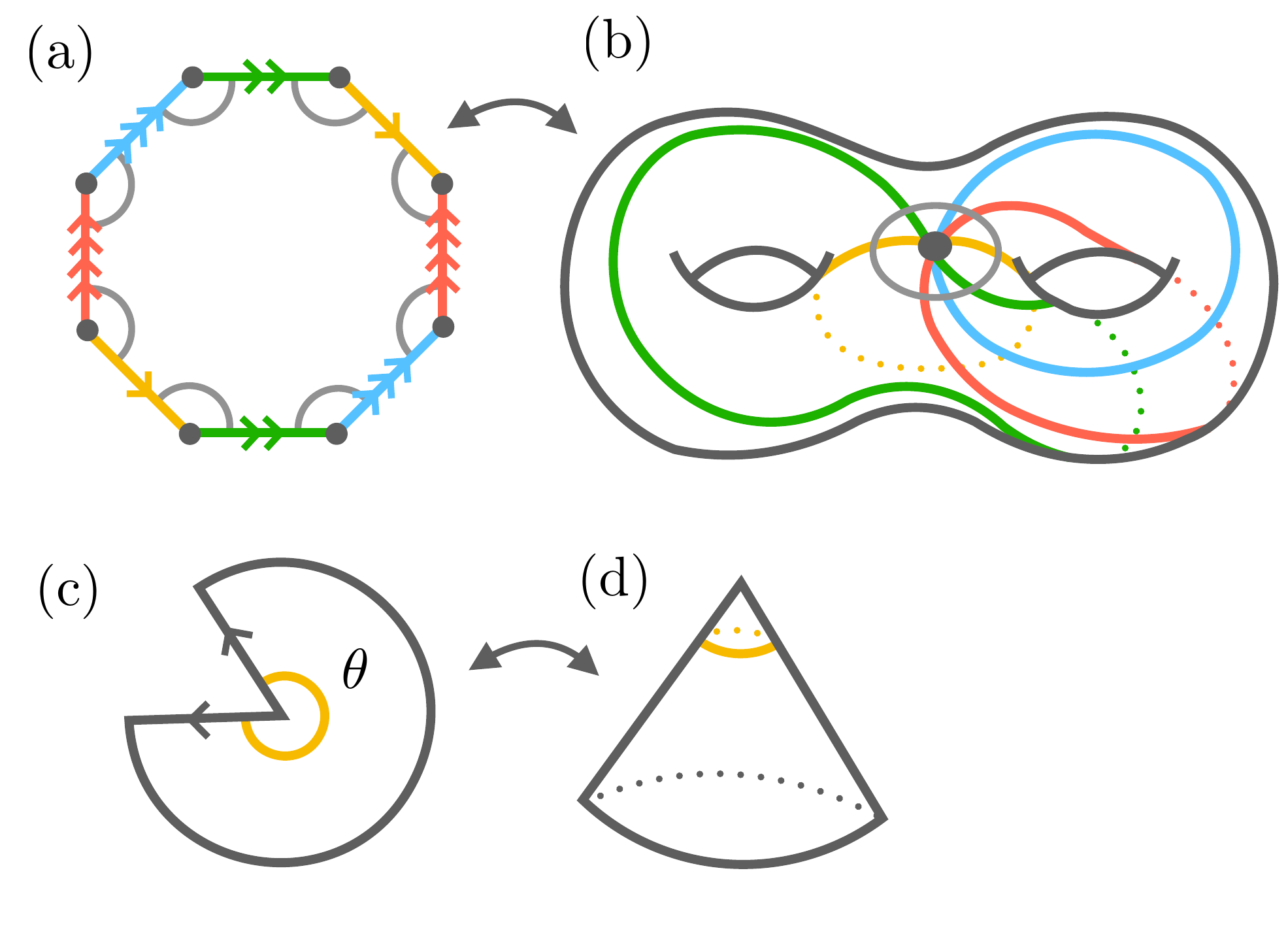}%
		\caption{\label{fig:torus} \textbf{Pseudo-Brillouin-zone topology and cone angle $ \bm{\theta} $ for a point of singular curvature}. (\textbf{a}) 
		The PBZ for a quasicrystal with 8-fold rotational symmetry. Opposite edges of a PBZ are identified. (\textbf{b}) Analogous to the topological equivalence between a regular Brillouin zone and a torus, a PBZ is topologically equivalent to a higher genus torus, in this example, the genus is 2.  (\textbf{c-d}) Illustration of the `cone angle', $ \theta $, that characterises points of singular curvature. In this example, $ \theta < 2\pi $, and therefore the apex of the cone has a positive singular curvature.
		}
	\end{figure}
	
	The total Gaussian curvature, $ \kappa $, for a manifold with, $ g>1 $, is negative, as seen by Gauss-Bonnet~\cite{lee2006riemannian},
	\begin{equation}
	\chi = \frac{1}{2\pi} \int_{\mathcal{M}} \kappa \ dS
	\end{equation}
	where $ dS $ is an element of area on the manifold $ \mathcal{M} $. For the PBZ this curvature is located entirely at the corners---since this surface is otherwise flat---where it is singular. We can therefore decompose Gauss-Bonnet into a sum over singular contributions,
	\begin{equation}
	\int_{\mathcal{M}} \kappa \ dS = \sum_{i \, \in \, \text{corners}} \kappa_i.
	\end{equation}
	Moreover, a singular curvature (in an otherwise flat surface) can be related to a `cone angle' $ \theta $~\cite{fillastre2017remark},
	\begin{equation}
	\kappa_i = 2\pi - \theta_i
	\end{equation}
	with this illustrated in Fig.~\ref{fig:torus}c-d for $ \theta < 2\pi $. In the current context, we have $ \theta_i = 2\pi N_\text{t} $, where $ N_\text{t} $ is the turning number associated to a corner. We therefore arrive at the result,
	\begin{equation}
	\chi = \xi (1-N_\text{t})
	\end{equation}
	where $ \xi $ is the number of vertices---$ \xi = 1 $ for even $ m $ and $ \xi = 2 $ for odd $ m $. This completes the connection between the PBZ genus and turning number,
	\begin{align}
	N_\text{t}^\text{even} &= 2g-1, \\
	N_\text{t}^\text{odd} &= g.
	\end{align}
	We see from this reasoning that the PBZ topology manifests as points of singular curvature which in turn directly determines the turning number for trajectories around these special points.

	\section{Identifying Nontrivial Turning Numbers from Magnetic Breakdown}

	Here we derive a general statement concerning the relationship between magnetic breakdown transitions and the presence of nontrivial turning numbers. That is: given a magnetic breakdown transition that occurs at fixed Fermi energy, the sum of all turning numbers must be conserved, which implies that \emph{at least one} Fermi surface contour has nontrivial turning number ($ |N_\text{t}| \neq 1 $) if there is a change in the number of frequencies modulo 2. 
	
	We first highlight that magnetic breakdown occurs in two varieties, referred to as interband and intraband. As their names suggest, interband occurs between two bands, typically with an avoided crossing, whereas intraband occurs in a single band, typically at a saddle point. Both types can be described by a two-in-two-out breakdown vertex, as shown in Fig.~\ref{fig:SuppConservationTurning}, in which case they simply label the two unique orientations of incoming and outgoing vertices (up to rotations). 
	
	The associated scattering matrices that connect incoming and outgoing edges are as follows. For interband one has (using the notation in Fig.~\ref{fig:SuppConservationTurning}),
	\begin{equation}\label{eq:interc}
	\begin{pmatrix}
	c_1^- \\ c_2^-
	\end{pmatrix}
	= \cS_\text{interband} 
	\begin{pmatrix}
	c_1^+ \\ c_2^+
	\end{pmatrix}
	\end{equation}
	with scattering matrix~\cite{phases},
	\begin{align}
		\cS_\text{interband} = \begin{pmatrix}
			\tau \e^{\ii (\omega+\theta)} &  \rho  \\
			\rho & - \tau \e^{-\ii (\omega+\theta)} 
		\end{pmatrix}, \label{eq:intermatrix}
	\end{align}
	where $ \rho = e^{-\pi \mu} $, $ \tau \equiv \sqrt{1-\rho^2} $, $ \omega = \mu - \mu \ln \mu + \arg\Gamma(\ii\mu) +\pi/4 $, $ \theta $ is the phase of the matrix element that opened the avoided crossing, and $ \mu = \frac{1}{2} ab \ell_B^2 $ with $ a $ and $ b $ the hyperbolic axes of the avoided crossing. Whereas for intraband one has,
	\begin{equation}\label{eq:intrac}
	\begin{pmatrix}
	c_\nwarrow \\ c_\searrow
	\end{pmatrix}
	= \cS_\text{intraband} 
	\begin{pmatrix}
	c_\nearrow \\ c_\swarrow
	\end{pmatrix}
	\end{equation}
	with scattering matrix,
	\begin{equation}
	\cS_\text{intraband} =  
	\begin{pmatrix}
	\mathcal{T} &  \mathcal{R}  \\
	\mathcal{R} & \mathcal{T} 
	\end{pmatrix}
	\end{equation}
	where,
	\begin{equation}
	\mathcal{T}(\mu') = \e^{i\phi(\mu')} \frac{\e^{\pi \mu'/2}}{\sqrt{2\cosh(\pi\mu')}}
	\end{equation}
	with $ \mathcal{R}(\mu') = - i \e^{\pi\mu'} \mathcal{T}(\mu') $, $ \phi(\mu') = \arg[\Gamma(\frac{1}{2}-i\mu')]+\mu' \log |\mu'| - \mu' $, and $ \mu' = \text{sgn}(E) \frac{1}{2} ab \ell_B^2 $, with $ ab $ as for interband and $ E $ is the energy relative to the crossing point of the saddle point.
	
	\begin{figure}[t]
		\centering
		\includegraphics[trim={.1cm 5.5cm 5.cm .2cm}, clip,width=.99\linewidth]{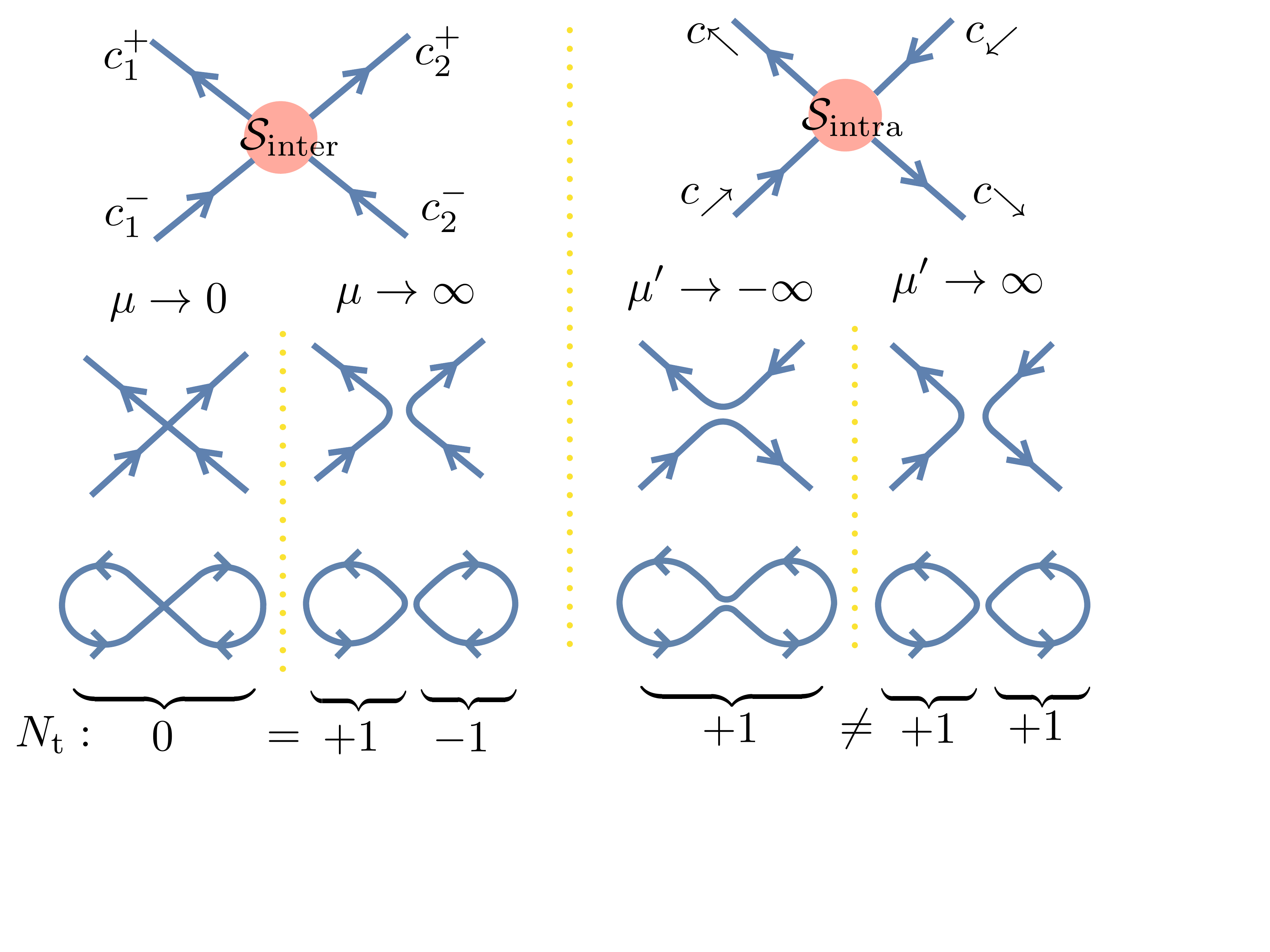}%
		\caption{\label{fig:SuppConservationTurning} \textbf{The two types of magnetic breakdown, their limiting behaviour and conservation of turning number}. (\textbf{Top}) 
			Diagrams of two-in-two-out breakdown vertices for interband (left) and intrabrand (right). (\textbf{Middle}) Sketch of the limiting behaviour of the scattering matrices---highlighting that a change in Fermi energy (sign of $ \mu' $) is required to change the Fermi surface connectivity for intraband.  (\textbf{Bottom}) Examples of magnetic breakdown transitions, demonstrating that the total turning number is conserved for interband but is not conserved for intraband.
		}
	\end{figure}
	
	A change in the Fermi surface connectivity occurs when the scattering matrix transitions between diagonal and off-diagonal. Crucially, this can occur at fixed Fermi energy for interband. Whereas for intraband, the Fermi energy is required to cross the saddle point. For interband one has,
	\begin{align}
		\mu \to \infty,& \qquad \cS_\text{interband} \to \begin{pmatrix}
			e^{i\theta} & 0 \\
			0 & -e^{i\theta}
		\end{pmatrix} \\
		\mu \to 0,& \qquad \cS_\text{interband} \to \begin{pmatrix}
			0 & 1 \\
			1 & 0
		\end{pmatrix}.
	\end{align}
	Whereas for intraband one has,
	\begin{align}
		\mu' \to \infty,& \qquad \cS_\text{intraband} \to \begin{pmatrix}
			1 & 0 \\
			0 & 1
		\end{pmatrix} \\
		\mu' \to 0,& \qquad \cS_\text{intraband} \to \begin{pmatrix}
			1/\sqrt{2} & -i/\sqrt{2} \\
			-i/\sqrt{2} & 1/\sqrt{2}
		\end{pmatrix}\\
		\mu' \to \infty,& \qquad \cS_\text{intraband} \to \begin{pmatrix}
			0 & -i \\
			-i & 0
		\end{pmatrix}.
	\end{align}
	One can therefore distinguish the type of magnetic breakdown by assessing whether or not magnetic breakdown occurs at fixed Fermi energy.

	Furthermore, the sum of all turning numbers is conserved across an interband transition. Denoting the, $ n $, turning numbers before the transition, $ N_\text{t}^i $, and the, $ n' $, turning numbers after by, $ {N_\text{t}'}^i $, one has,
	\begin{equation}\label{eq:tcons}
	\sum_{i=1}^{n} N_\text{t}^i = \sum_{i=1}^{n'} {N_\text{t}'}^i
	\end{equation}
	An example is shown in Fig.~\ref{fig:SuppConservationTurning}, in which a figure of eight curve with a nontrivial turning number of zero is split into two trivial curves with turning numbers, $ \pm 1 $,  as $ \mu: 0 \to \infty $. For intraband transitions, the turning number is not conserved, with an example shown in Fig.~\ref{fig:SuppConservationTurning}.
	
	Given that the sum of all turning numbers is conserved, a statement about the presence of nontrivial turning numbers can be made using the following argument. First, we assume all turning numbers are trivial before and after a transition, $  |N_\text{t}^i| = |{N_\text{t}'}^i| = 1 $ for all $ i $. Then we have,
	\begin{equation}
	\sum_{i=1}^{n} N_\text{t}^i = n \quad \text{mod} \ 2,
	\end{equation}
	and,
	\begin{equation}
	\sum_{i=1}^{n} {N_\text{t}'}^i = n' \quad \text{mod} \ 2.
	\end{equation}
	Since the sum of all turning numbers is conserved \eqref{eq:tcons},
	\begin{equation}
	n = n' \quad \text{mod} \ 2.
	\end{equation}
	However this is not necessarily true, as in the case shown in Fig.~\ref{fig:SuppConservationTurning} We therefore have a contradiction: if, $ n \neq n' \ \text{mod} \ 2 $, the assumption that all turning numbers before and after are trivial must be incorrect. Therefore \emph{at least one} turning number must be nontrivial.
	
	More precisely, this test identifies that there is at least one \emph{even} turning number (nontrivial by definition). However, a nontrivial odd turning number may be present and would not be identified by this method. Therefore a change in parity of the total number of turning numbers is a sufficient but not necessary condition for the presence of a nontrivial turning number. 
	
	To summarize, if magnetic breakdown occurs at fixed Fermi energy, this must be due to an interband transition. In this case, the sum of all turning numbers is conserved. Furthermore, if the total number of turning numbers (or frequencies in the quantum oscillations) changes modulo 2, there must be \emph{at least one} nontrivial turning number.

	\begin{figure}[tbh]
		\centering
		\includegraphics[trim={.5cm 2.1cm 12.cm 2.2cm}, clip,width=.6\linewidth]{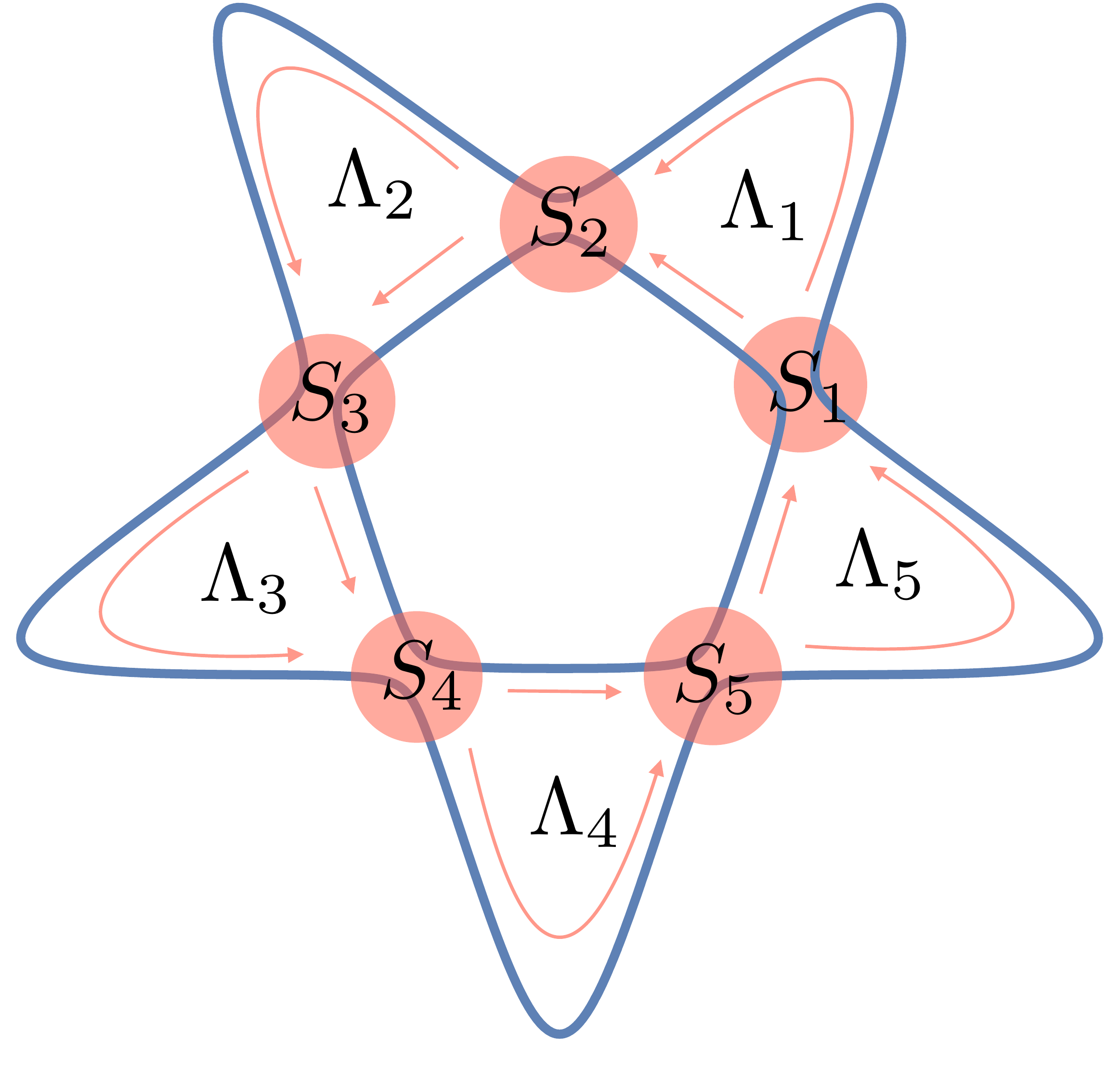}%
		\caption{\label{fig:SuppScatteringQuant} \textbf{Scattering matrix generalisation of semiclassical quantisation}. Sketch showing how a generalised semiclassical quantisation is applied to the spiral holonomy contours. Here the wavepacket undergoes partial magnetic breakdown at avoided crossings (pink disks) described by scattering matrices, $ S_i $, but follows classical trajectories along intermediate contours with phase evolution described by diagonal matrices, $ \Lambda_i $.
		}
	\end{figure}

	\section{Numerical Approach for Magnetic Breakdown Spectrum}
	
	In order to compute the quantum oscillation frequency spectrum in Fig.~\ref{fig:MBNFE}, we use a generalisation of semiclassical quantisation~\cite{slutskin1968dynamics,alexandradinata2018semiclassical}. In which we replace the self intersections (in the first order regime) of the Fermi surface contours with interband scattering matrices given in \eqref{eq:intermatrix}, and as shown in Fig.~\ref{fig:SuppScatteringQuant}. 
	A diagonal matrix accounts for phase evolution along intervening contours,
	\begin{align}
		\Lambda = \begin{pmatrix}
			\e^{\ii \Omega_1} & 0 \\
			0 & \e^{\ii \Omega_2} 
		\end{pmatrix},
	\end{align}
	where $ \Omega_i \equiv \ell_B^2 S_i +{\varphi_\text{M}}_i + {\varphi_\text{B}}_i $, is the sum of area, Maslov and Berry contributions. While the total unitary evolution is given by the product
	\begin{align}
		U \equiv \prod_{j=1}^{5} \Lambda_j \cS_j.
	\end{align}
	By requiring single-valuedness, $ U \psi = \psi $, an equation that generalises semiclassical quantisation for intermediate breakdown is provided by,
	\begin{align}
		\text{det}(U-I) = 0. \label{eq:det}
	\end{align}
	This equation recovers the usual semiclassical quantisation \eqref{eq:onsager} in both limits $ P_\text{MB} \to 0,\,1 $.
	
	The spectrum is then computed by constructing the density of states at the Fermi energy as a function of $ 1/B $. Here this consists of delta peaks centred at each solution in $ B $ to \eqref{eq:det}, as shown in Fig.~\ref{fig:SuppSpectrum}a. The discrete Fourier transform~\cite{apodization} is then taken over a finite range of $ 1/B $ (a small non-zero broadening is used for the delta peaks which does not affect the results). The size of this range is chosen so that it is large enough to provide sufficient resolution but small enough to distinguish the change in frequencies across the transition. All frequencies present in the Fourier transform can be labelled using the semiclassical trajectories shown in Fig.~\ref{fig:SuppSpectrum}b.

	\begin{figure}[t]
		\centering
		\includegraphics[trim={1.cm .1cm 13.1cm .2cm}, clip,width=.85\linewidth]{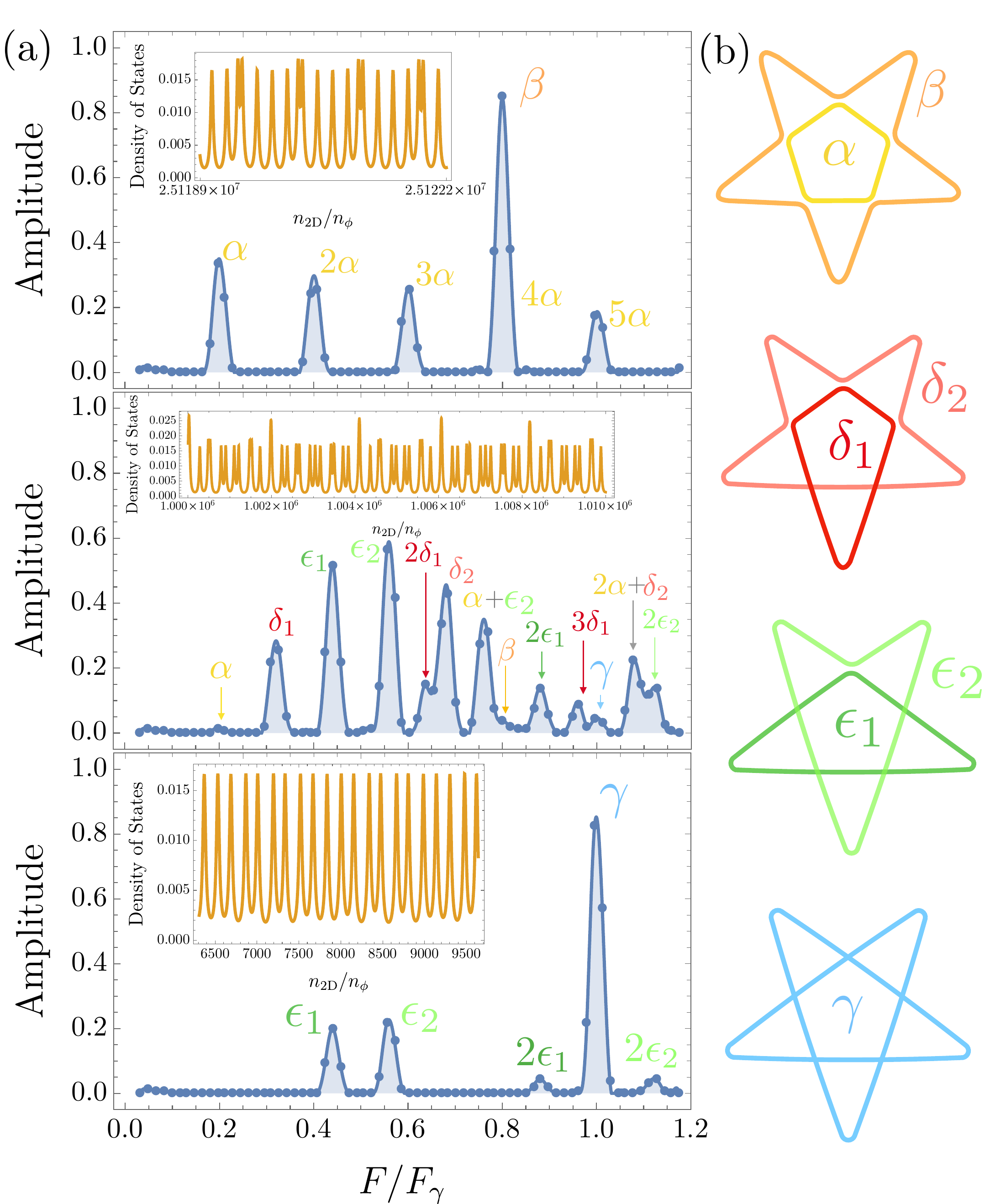}%
		\caption{\label{fig:SuppSpectrum} \textbf{Examples of frequency spectrum for three regimes of magnetic field}. (\textbf{a}) Fourier transform of density of states at fixed Fermi energy (shown inset). These a slices of the density plot in Fig.~\ref{fig:MBNFE} at the field values (top to bottom): $ n_\phi/n_\text{2D} = 10^{-4} $, $ n_\phi/n_\text{2D} = 10^{-6} $ and $ n_\phi/n_\text{2D} = 10^{-7} $. (\textbf{b}) Semiclassical contours used to label frequency spectrum plots in (a).
		}
	\end{figure}

\section{Stability of perturbation theory}

We provide a simple argument for why perturbation theory is stable for the models studied in the main text. Very general arguments are known for one-dimensional tight binding models with sufficiently well behaved quasiperiodic potential~\cite{dinaburg1975one,steinhardt1987physics}. Nearly-free-electron or weakly coupled tight binding models have received less rigorous arguments~\cite{lu1987electronic}. We therefore provide a clear and complete argument in the following to address any possible concerns. Specifically, we wish to show that there are no singular features in the perturbative limit. 

The argument we use hinges on the number theoretic properties of the irrational number, $ \xi $, that underlies the incommensurability of the particular model. Namely we use the convergence properties of rational approximations to Diophantine numbers to avoid the so called `small denominator problem'. The same properties are central to almost all arguments on the validity of perturbation theory for quasiperiodic systems~\cite{dinaburg1975one,steinhardt1987physics}. Moreover it is the same property that is central in the well known KAM theory~\cite{lowenstein2012essentials}.

%

{\emph{Model}.---} We focus on a simple one-dimensional version of the nearly-free-electron quasicrystal discussed in our paper, that will highlight the key aspects here. Consider the Hamiltonian
\begin{align}
\hH \equiv \frac{\hp^2}{2m}+\hV, \label{eq:hamiltonian}
\end{align}
where
\begin{align}
\hV \equiv  V_0 \cos \kappa \hx  + V_0\cos \xi \kappa \hx, \label{eq:potential}
\end{align}
with $ \xi $ irrational. Any eigenstate of \eqref{eq:hamiltonian} can be written in a momentum space basis as
\begin{align}
|\psi_k \ra  = \sum_{G} a^k_G |k-G\ra, \label{eq:eigstate}
\end{align}
where the sum runs over all combinations, $ G  = iG_1+jG_2$, of the two basic momentum transfers, $ G_1 \equiv \kappa $ and $ G_2 \equiv \xi \kappa $. 

For perturbation theory to be valid one requires that the eigenstates and energies, $ |\psi_k\ra $ and $ E_k $, of the full Hamiltonian, $ \hH $, remain arbitrarily close to those of the unperturbed Hamiltonian, $ \hH_0 = \hp^2/2m $. That is,
\begin{align}
|\psi_k\ra \to |k\ra, \quad E_k \to \epsilon_k \quad \text{for} \quad \frac{V_0}{\epsilon_\kappa} \to 0
\end{align}
where $ \epsilon_k \equiv \hbar^2k^2/2m $ and $ \epsilon_\kappa $ is a characteristic energy scale.

The potential issue of perturbation theory is due to set of momentum transfers, $ G $, in \eqref{eq:eigstate}, being dense---which is due to the irrationality of $ \xi $. This means that any free particle state, $ |k\ra $, is essentially resonant with a large number of states, $ |k-G\ra $,
\begin{align}
\epsilon_k \simeq \epsilon_{k-G}. \label{eq:resonance}
\end{align}
Moreover, the potential $ \hV $ will effectively couple these with a matrix element, $ V_0^\text{eff} $, such that within the subspace of $ |k\ra $ and $ |k-G\ra $, one will have an effective Hamiltonian,
\begin{align}
\hH^\text{eff} = \begin{pmatrix}
\epsilon_k & V_0^\text{eff}\\
V_0^\text{eff} & \epsilon_{k-G}
\end{pmatrix}.\label{eq:Heff}
\end{align}
The relative sizes of $ \epsilon_k-\epsilon_{k-G} $ and $ V_0^\text{eff} $ are then crucial to determining whether or not the two states are strongly mixed, and therefore whether perturbation theory is valid. Specifically, if
\begin{align}
\frac{V_0^\text{eff}}{\epsilon_k-\epsilon_{k-G}} \ll  1 \label{eq:condition}
\end{align}
is satisfied, the mixing is weak, and perturbation theory is valid. 


{\emph{Diophantine approximation}.---} A bound can be found for the energy difference, $ \epsilon_k-\epsilon_{k-G} $, by using standard results from number theory. In particular, we wish to find the dependence of $ \epsilon_k-\epsilon_{k-G} $ on the total number of momentum transfers, 
\begin{align}
n \equiv |i| + |j|.
\end{align}

This connection is made by first assuming that $ |k-G\ra $ is not exactly degenerate with $ |k\ra $. We then find two approximate equations to be satisfied by $ G $, in order to satisfy \eqref{eq:resonance},
\begin{align}
i + j \xi  &\simeq 0, \label{eq:ij1}\\
i + j \xi  &\simeq 2k/\kappa. \label{eq:ij2}
\end{align}
Since $ i + j \xi $ is dense over the real numbers, the convergence properties the second equation are identical to  first, we therefore only discuss the first. The connection to diophantine approximation is clear since \eqref{eq:ij1} is equivalent to,
\begin{align}
\xi \simeq -\frac{i}{j}.
\end{align}
The ratios of $ i $ and $ j $ are therefore successive rational approximants to $ \xi $.

The convergence properties of \eqref{eq:ij1} only depend on the degree of irrationality of $ \xi $. This is characterised by the irrationality measure $ \mu(\xi) $, defined by the following: $ \mu $ is the smallest number such that the inequality~\cite{bugeaud2012distribution}
\begin{align}
\left|i+j\xi\right| > \frac{1}{j^{\mu-1+\delta}}
\end{align}
holds for any $ \delta>0 $ and all integers $ i $ and $ j $, with $ j $ sufficiently large. It is known that $ \mu \ge 2 $ for all irrationals, and moreover $ \mu = 2 $ for almost all real numbers, with respect to the Lebesque measure. One therefore has that for large $ j $,
\begin{align}
\left|i+j\xi\right| \sim \frac{1}{j},
\end{align}
for almost all real numbers $ \xi $. A notable exception to this is for so called Liouville numbers, for which $ \mu = \infty $, which therefore have convergence that is faster than any power law. However, these numbers have zero measure, and are therefore not a problem in any realistic situation. All irrational numbers that are not Liouvillian are referred to as Diophantine.

Finally, since $ n \sim j $, the above result gives
\begin{align}
\epsilon_k-\epsilon_{k-G} \sim \frac{\epsilon_\kappa}{n} .\label{eq:convergence}
\end{align}
Therefore, the closest approach in energy of two states $ |k\ra $ and $ |k-G\ra $ is given by a power law of the total number of momentum transfers, $ n $, connecting the two states. In other words, although a state $ |k-G\ra $ can be arbitrarily close to degeneracy with $ |k\ra $, it requires going to high orders of momentum transfers.


{\emph{Effective Hamiltonian theory}.---} Before addressing $ V_0^\text{eff} $, we explain more precisely what we mean by an \emph{effective} coupling from the potential $ \hV $. Clearly the matrix element of $ \hV $ between the two states is zero for $ G $ with $ n > 1 $,
\begin{align}
\la k-G | \hV | k \ra = 0, \quad n>1.
\end{align}
Instead, $ V_0^\text{eff} $, is constructed within the `effective Hamiltonian theory' or `partitioning technique'~\cite{suzuki1983degenerate}.
Here one derives an effective interaction, $ \hV^\text{eff} $, that acts in a given subspace defined by the projector, $ \hP $, such that,
\begin{align}
\hH^\text{eff} = \hP \hH \hP + \hV^\text{eff},
\end{align}
has eigenvalues agreeing with the full Hamiltonian $ \hH $. Here we have,
\begin{align}
\hP = | k \ra \la k | + | k-G \ra \la k-G |,
\end{align} 
therefore,
\begin{align}
\hP \hH \hP &= \epsilon_k| k \ra \la k | + \epsilon_{k-G}| k-G \ra \la k-G |,\\
\hV^\text{eff} &= V_0^\text{eff} | k \ra \la k-G | + V_0^\text{eff} | k-G \ra \la k |,
\end{align}
as given in the above. 

The particular value of $ V_0^\text{eff} $ is then given by a perturbative expansion. Intuitively, at $ m $'th order, this is a sum over all processes that involve $ m $ momentum transfers. Since the two states require a minimum of $ n $ momentum transfers to be connected, the lowest order of this expansion will have the form,
\begin{align}
V_0^\text{eff} \sim g(n)\frac{V_0^n}{\epsilon_\kappa^{n-1}},
\end{align}
where $ g(n) $ counts the all possible trajectories between $ |k\ra $ and $ |k-G\ra $. Since $ G $ is defined on a two-dimensional grid, 
\begin{align}
g(n) &= \begin{pmatrix}
(|i|-1) + (|j|-1)\\
|i|-1,|j|-1
\end{pmatrix}\\ 
&= \frac{(|i|-1) + (|j|-1))!}{(|i|-1)!(|j|-1)!}\\
&\sim 2^n
\end{align}
where the last expression is the strongest possible scaling. Therefore the scaling remains exponential in $ n $,
\begin{align}
V_0^\text{eff} \sim \frac{(2V_0)^n}{\epsilon_\kappa^{n-1}},
\end{align} 
and rescaling, $ V_0 \to V_0/2 $, removes this additional factor,
\begin{align}
V_0^\text{eff} \sim \frac{V_0^n}{\epsilon_\kappa^{n-1}}. \label{eq:v0eff}
\end{align} 

Using \eqref{eq:v0eff} alongside the convergence property of diophantine numbers \eqref{eq:convergence}, one finds for the ratio in \eqref{eq:condition},
\begin{align}
\left(\frac{V_0}{\epsilon_\kappa}\right)^n n \ll 1,
\end{align}
which holds for all $ n $, with $ V_0/\epsilon_\kappa $ sufficiently small. Therefore, despite the power law convergence of the energy difference, the coupling decays exponentially, and can therefore always be made sufficiently small to ensure weak mixing between these states.

The above result is valid for near degenerate states, however it is trivial to extend to the case of a degeneracy: In such a situation, the energy shift given by the effective Hamiltonian in \eqref{eq:Heff} is simply,
\begin{align}
\frac{\Delta E^\text{deg}}{\epsilon_\kappa }\equiv \frac{E_k^\text{deg} - \epsilon_k}{\epsilon_\kappa} \sim \left(\frac{V_0}{\epsilon_\kappa}\right)^n n,
\end{align}
which can again be made arbitrarily small. Additionally, all other states are then necessarily non-degenerate with $ |k\ra $, and as such their respective energy shifts can also be made arbitrarily small.

An objection can be made to these results in that while the mixing with a single state is negligible, the state $ |k\ra $ is near degenerate with an exponentially large number of states---the net effect of which could cause perturbation theory to break down. Indeed, the total number of states at order $ n $, is given by,
\begin{align}
N(n) \simeq 4^n,
\end{align}
however this would simply alter the energy shift due to a single resonant state,
\begin{align}
\frac{\Delta E}{\epsilon_\kappa } \sim \frac{1}{2}\left(\left(\frac{V_0}{\epsilon_\kappa}\right)^n n\right)^2,
\end{align}
by a multiplicative factor,
\begin{align}
\frac{\Delta E}{\epsilon_\kappa } \sim \frac{1}{2}\left(\left(\frac{2V_0}{\epsilon_\kappa}\right)^n n\right)^2,
\end{align}
which can also be removed by rescaling, $ V_0 \to V_0/2 $. 
Furthermore, the total energy shift due to resonances at all orders $ n $ is well controlled since,
\begin{align}
\frac{\Delta E_\text{tot}}{\epsilon_\kappa } \sim \sum_{n=n_0\gg1}^{\infty} \left(\frac{V_0}{\epsilon_\kappa}\right)^n \sim \frac{V_0^{n_0}}{1-V_0} \sim V_0^{n_0}.
\end{align}

{\emph{Generalizations}.---} Our arguments can be trivially generalised to higher dimensional quasiperiodic systems: The resonant states are those nearby the Ewald sphere, all possible trajectories $ g(n) $ scales as $ d^n $, (where $ d $ is the number of linearly independent basis vectors for the diffraction pattern), and the total number of states at order $ n $ scales as $ (2d)^n $. However all of these do not change the conclusions found here.

Moreover one can generalise to potentials, $ \hV' $, with non-zero matrix elements between all states. With the additional assumption that these matrix elements also decay exponentially,
\begin{align}
\la k-G | \hV' | k \ra \lesssim \left(\frac{V_0}{\epsilon_\kappa}\right)^n.
\end{align}
This dependence is reasonable for the following reason: The underlying ionic potential of the quasicrystal might have large Fourier components with arbitrary small wavevector, however the electronic density will naturally screen any small wavevector, long wavelength fluctuations. The resulting electronic density---which is essentially what $ \hV $ encodes---will have exponentially weak Fourier components for small wavevectors. 

Furthermore, while these results are relevant to the nearly-free-electron model we study, it is simple to extend these to the case of \ang{30} twisted bilayer graphene (or in general to weakly coupled tight binding layers). 

As with the nearly-free-electron case, we expect perturbation theory to be unstable due to resonances between the unperturbed eigenstates. Here this is between the eigenstates, $ |\kk,n\ra $ and $ |\tkk, \tn \ra $, of each layers tight binding Hamiltonian, $ H_\parallel $ and $ \tH_\parallel $. (Instead of free particle states, as in the nearly-free-electron case.)

For a particular state, $ |\kk,n\ra $, on layer 1, the resonance condition with the state, $ |\tkk, \tn \ra $, on layer 2, is given by, 
\begin{align}
E(\kk) \simeq \tE(\tkk), \quad \kk+\GG = \tkk +\tGG, \label{eq:tblgcondition}
\end{align} 
where $ E(\kk) $ and $ \tE(\tkk) $ are the bandstructures of the unperturbed layers 1 and 2. By rewriting this as,
\begin{align}
E(\kk) \simeq \tE(\kk + \GG - \tGG),
\end{align}
and defining the set of vectors $ \{\GG_{\kk}\} $ such that, $ E(\kk) = \tE(\kk + \GG_{\kk}) $. We see that in order to satisfy \eqref{eq:tblgcondition} given a particular $ \kk $, we must find $ \GG-\tGG $ such that,
\begin{align}
\GG-\tGG \simeq \GG_{\kk}.
\end{align}
for an arbitrary element of $ \{\GG_{\kk}\} $.
This is again a problem of rational approximation, and therefore an approximate solution to \eqref{eq:tblgcondition} will satisfy,
\begin{align}
E(\kk) - \tE(\tkk) \sim \frac{t}{n}
\end{align}
where $ n $ is the total number of momentum transfers.


The next step in our argument is simpler in this case, since the matrix element between two resonant states is in general non-zero, and therefore it is unnecessary to consider an effective coupling (for $ t_\perp/t \ll 1 $). This matrix element is approximately given by,
\begin{align}
\la \tkk, \tn | H_\perp | \kk, n \ra \sim t_\perp(\kk+\GG).
\end{align}
Since, $ |\GG| \sim n \gg 1 $, one has, $ | \kk + \GG | \sim n $, and therefore,
\begin{align}
\la \tkk, \tn | H_\perp | \kk, n \ra \sim t_\perp \e^{-n},
\end{align}
because, $ t_\perp(\kk) $, decays exponentially for large $ |\kk| $.

Finally, the ratio that controls the mixing between these states is therefore given by,
\begin{align}
\frac{\la \tkk, \tn | H_\perp | \kk, n \ra}{E(\kk) - \tE(\tkk)} \sim \frac{t_\perp}{t} \e^{-n} n \ll 1.
\end{align}
Additionally, there will be effective intra-layer couplings between, $ | \kk, n \ra $ and $ | \kk + \tGG, n \ra $, however this is at second order, $ t_\perp^2/t $, so is negligible compared to the first order, inter-layer, mixing.

{\emph{Discussion}.---} Overall we have shown that due to generic limitations on rational approximations of irrational numbers, resonances appear as a power law in $ n $. While the coupling between these is exponentially small in $ n $. Therefore the mixing between these resonant states remains well controlled in the perturbative limit. 

Essentially, we have demonstrated that although the actual spectrum/dispersion is complicated---with the possibility of gaps of arbitrarily small size---there is a sense in which the spectrum/dispersion remains simple. That is, if one looks at the spectrum/dispersion with a non-zero energy resolution, any gaps below this resolution can be essentially ignored, leaving those above this resolution. This structure is precisely what is observed in the numerous ARPES studies on quasicrystals. There one can effectively see a continuous dispersion with some large gaps at the Fermi energy. While there are necessarily gaps in this seemingly continuous dispersion, the limited energy resolution washes these out. Precisely the same idea is central throughout our work, with the energy resolution being set by the cyclotron frequency, $ \hbar \omega_c $, allowing the physics of quantum oscillations to be determined solely by the larger gaps. 
	
	\vspace{200em}

\end{twocolumngrid}

\end{document}